\documentclass[pra,aps,twocolumn,floatfix,10pt]{revtex4-1}

\usepackage{graphicx} 
\usepackage{amsmath}
\usepackage{hyperref}

\begin{document} 
\title{Non-perturbative method to compute thermal correlations in one-dimensional systems}   

\author{Stefan Beck$^{1,2}$, Igor E. Mazets$^{1,2}$, and Thomas Schweigler$^1$} 
\affiliation{
$^1$\, Vienna Center for Quantum Science and Technology, Atominstitut, TU~Wien,~Stadionallee~2,~1020~Vienna,~Austria \\
$^2$\, Wolfgang Pauli Institute c/o Fakult\"{a}t f\"{u}r Mathematik,
Universit\"{a}t Wien, Oskar-Morgenstern-Platz 1, 1090 Vienna, Austria}

\begin{abstract} 
We develop a highly efficient method to numerically simulate thermal fluctuations and correlations in non-relativistic continuous bosonic
one-dimensional systems. 
The method is suitable for arbitrary local interactions as long as the system remains dynamically stable.
We start by proving the equivalence of describing the systems through the transfer matrix formalism and a Fokker-Planck equation for a distribution evolving in space. 
The Fokker-Planck equation is known to be equivalent to a 
stochastic differential (It\={o}) equation. 
The latter  is very suitable for computer simulations, allowing the calculation of  any desired correlation function. 
As an illustration, we apply our method to the case of two tunnel-coupled quasi-condensates of bosonic atoms.
The results are compared to the predictions of the sine-Gordon model for which we develop analytic expressions directly from the transfer matrix formalism.
\end{abstract}
\maketitle 

\section{Introduction} 
\label{intr} 

One-dimensional (1D) quantum systems attract much attention because their dynamics is strongly affected by the restricted phase space available for scattering~\cite{Giamarchi04,Cazalilla11,YUROVSKY200861}.
As a result, such systems may exhibit a number of non-trivial properties. For example, one can observe integrability as well as its breakdown. 
The most well-known 1D theoretical models include the Lieb-Liniger model~\cite{Lieb63,Lieb63b}, the quantum Luttinger liquid theory~\cite{Luttinger63}, and the sine-Gordon model~\cite{cuevas2014sine,Coleman75,Mandelstam}.
 
Experimentally, 1D quantum systems are realized by freezing the motion of particles in two tightly confined dimensions.
Available 1D systems range from ultracold atomic gases~\cite{bloch2008many,proukakis2017universal} to slow-light polaritons~\cite{chang2008} as well as superfluid $^4$He atoms adsorbed in nanometer-wide pores~\cite{Wada01}. 
Several recent experimental studies underline the importance of 1D systems as a testbed for theoretical ideas, in and out of equilibrium~\cite{Langen2015,Bordia2017,Rauer17,Schweigler17}.  

The rapid progress in the preparation, manipulation and characterization of experimental systems, especially in the realm of ultracold atoms, also leads to a need for ever improving theoretical descriptions. In particular, the recent measurement of higher-order correlation functions in 1D quasi-condensates~\cite{Schweigler17} calls for novel theoretical methods beyond the perturbative approach.

This paper presents a detailed description of a versatile, non-perturbative method to calculate thermal correlations of multicomponent bosonic fields in the mean-field approximation in 1D. 
The search for the method has been inspired by the observation that thermal Gaussian fluctuations in a 1D system of bosons with quadratic effective Hamiltonian can be 
described by the Ornstein-Uhlenbeck stochastic process~\cite{stimming2010fluctuations}, where the spatial co-ordinate plays the 
role usually taken by time in conventional applications of random processes. 
The method developed in Ref.~\cite{stimming2010fluctuations} has been  applied to the analysis of experimental data 
for 1D systems~\cite{twinbeams,Gring2012} 
and, in its extended version~\cite{OU2D}, even for two-dimensional systems~\cite{Seo2014}. 

These works relied on the absence of interactions between quasiparticles in the theoretical description, i.e., quadratic Hamiltonians. We extended our previous approach to systems where such interactions have to be taken into account, i.e., to non-quadratic Hamiltonians. It is surprising that the description via a stochastic process along the 1D co-ordinate is also possible in this case. 
The stochastic process provides highly efficient numerical sampling of classical fields with the statistics given by the thermal equilibrium. We present the method and apply it to the case of two tunnel-coupled 1D quasi-condensates~\cite{Goldstein97,Whitlock03}.

We consider a system of bosonic particles effectively confined to a 1D geometry. On the quantum level, the particles are described by field operators obeying standard bosonic commutation rules~\cite{Giamarchi04,Cazalilla11,YUROVSKY200861}. Different components $\hat \psi _j$ of the field may correspond, for example, to atoms of the same kind trapped in different atomic waveguides (this case will be considered in detail in Sec.~\ref{model}). Another example would be the case of different atomic species or isotopes, or of different spin states of the same isotope trapped in the same waveguide. However, in the experimentally accessible range of temperatures and densities the equilibrium properties of ultracold gases of bosonic atoms in 1D are often dominated by classical thermal fluctuations~\cite{stimming2010fluctuations}. Therefore we may describe the system within the mean-field approximation, replacing the operators $\hat \psi _j$ with classical complex fields $\psi _j$. 

Let us now introduce the notation used in the paper. The co-ordinate in 1D is denoted by $z$. 
Instead of the ${\cal M}/2$ complex, $z$-dependent components $\psi _j$ (${\cal M}$ is an even integer number) we can also use ${\cal M}$ real components 
\begin{align} 
&q_{2j-1}(z)=\mathrm{Re}\, \psi _j(z), \qquad  
q_{2j}(z) = \mathrm{Im}\, \psi _j(z), ~~\notag \\ &\qquad \qquad 
\qquad \qquad \qquad \qquad j=1,\, 2, \, \dots \, ,\, {\cal M}/2.
\label{qj}
\end{align} 
In the following we will not indicate the $z$-dependence of the fields explicitly.
Without loss of generality, we assume that all the components are characterized by the same mass $m$~\footnote{Since we do not consider the time dynamics, but deal with thermal equilibrium properties only, the case of components with different masses $m_j$ can be reduced to the equal-mass case by simply rescaling the classical complex fields $\psi _j =\sqrt{m_j/m}\psi _j^\prime $.}. The Hamiltonian 
function of the system is then
\begin{align} 
H =&\int _{-L/2}^{L/2}
dz\, \Bigg{\{ }\sum _{j=1}^{\cal M} \left[ \frac {\hbar ^2}{2m} \left( \frac {\partial q_j}{\partial z}\right)^2-\mu _j q_j^2
\right]  \nonumber \\ & +  V (q_1, \, \dots \, ,\, q_{\cal M}) \Bigg{ \} }.
\label{hm1}
\end{align} 
Here $\mu _j$ is the chemical potential for the $j$th component. 
Since $q_{2j-1}$ and $q_{2j}$ are the real and imaginary part of the same complex field, there are only ${\cal M}/2$ independent chemical potentials. 
$V (q_1, \, \dots \, ,\, q_{\cal M})$ represents the local interaction energy density, which does not explicitly depend on $z$ (homogeneous system). 

We consider the thermodynamic limit $L\rightarrow \infty $, while the mass density remains constant. 
We therefore don't have to specify boundary conditions, as will become clear later.
In the following, we will refer to the integrand in Eq.~(\ref{hm1}) as energy density $h(q_1\, \, \dots \, , q_{\cal M})$. 

The paper is organized as follows. In Section~\ref{trmxf} we recall the basics of the transfer matrix formalism. In Section~\ref{from} 
we demonstrate the equivalence of this formalism to the description via an It\={o} stochastic ordinary differential equation. 
The full mean-field model describing both the phase and density fluctuations in two tunnel-coupled quasicondensates 
is introduced in Section~\ref{model}. 
In the same Section we discuss the sine-Gordon (SG) model as an approximation describing  
phase fluctuations only.  Within the SG model, we derive 
analytic expressions for the second and fourth moments of the relative phase difference, see Section~\ref{sGr}. 
In Section~\ref{rid} we present the results for the full model obtained numerically using our computational method based on the 
stochastic process. We compare these results with those of the SG model and discuss their similarities and differences. Section~\ref{conclusion} is devoted to conclusive remarks.

\section{The transfer matrix formalism} 
\label{trmxf}

In this section, we briefly recapitulate some basics of the transfer matrix formalism~\cite{Scalapino72,Krumhansl75,Currie80}. 
We are interested in equal-time correlations of the observables ${\cal F}^{(i)}\vert _{z_i}=
{\cal F}^{(i)} [q_1(z_i), \, \dots \, ,\, q_{\cal M}(z_i)]$ 
measured at different points $z_i$, $i=1,\, 2, \, \dots \, ,\, l$ at  
equilibrium with temperature $T$. We consider the general case where all these 
selected observables ${\cal F}^{(i)}$ may be different.

In order to introduce the transfer operator, we will first start with the partition function. 
In classical field approximation, the partition function of the system can be expressed as the functional integral 
\begin{equation} 
{\cal Z} =\int {\cal D}q_1 \, \dots \int {\cal D}q_{\cal M} \, e ^{-\beta H} ,
\label{I.1} 
\end{equation} 
where $\beta =1/(k_BT)$.
Recall the definition of the functional integral as the limit of the statistical sum of a model on a grid with the 
spacing $\Delta z \rightarrow 0$. We consider the two adjacent grid points $z_0$ and $z_0+\Delta z$. 
The fields at this points are denoted by $\{ q_j^\prime \}  $ and $\{ q_j \}$, respectively. 
Then Eq.~(\ref{I.1}) can be rewritten as 
\begin{align} 
\begin{split}
{\cal Z} = \int d^{\cal M}q \ &{\cal Z}_>\vert _{z_0+\Delta z}  \ \\ 
&\int d^{\cal M}q^\prime \  
e^{-\beta \Delta z\, h_\mathrm{fd} (\{ q_j\} , \, \{ q_j^\prime \}  )} \ {\cal Z}_<\vert _{z_0}   , 
\label{I.2} 
\end{split}
\end{align} 
where $\int d^{\cal M}q $ stands for $\int _{-\infty }^\infty dq_1\, \dots \int _{-\infty }^\infty dq_{\cal M} $ and we have introduced the finite-difference representation of the energy density 
\begin{align} 
h_\mathrm{fd} (\{ q_j^\prime \} , \, \{ q_j\}  )=&
\Bigg{\{ }\sum _{j=1}^{\cal M} \left[ \frac {\hbar ^2}{2m} \left( \frac {q_j-q_j^\prime }{\Delta z}\right)^2-\mu _j q_j^2
\right]  \nonumber \\ & +  V (q_1, \, \dots \, ,\, q_{\cal M}) \Bigg{ \} }.
\label{hfd}
\end{align}

The quantities ${\cal Z}_<\vert _{z_0} $ and ${\cal Z}_>\vert _{z_0+\Delta z} $ are the partition functions for the subregions $z < z_0$ and $z > z_0 + \Delta z $, respectively. We have 
\begin{equation} 
{\cal Z}_<\vert _{z_0}  =\int {\cal D}q_1 \, \dots \int {\cal D}q_{\cal M} \, e ^{-\beta \int _{-L/2}^{z_0}dz\, h}
\label{I.3} 
\end{equation} 
where the functional integral is only over fields in the region $[-L/2,z_0)$, not including the fields at the point $z_0$ itself. 
Analogously we define 
\begin{equation}
{\cal Z}_>\vert _{z_0 + \Delta z}  =\int {\cal D}q_1 \, \dots \int {\cal D}q_{\cal M} \, e ^{-\beta \int _{z_0 + \Delta z}^{L/2}dz\, h},
\end{equation}
where the functional integral is only over fields in the region $(z_0 + \Delta z,L/2]$, again not including the fields at the point $z_0 + \Delta z$ itself.
In Eq.~(\ref{I.2}), ${\cal Z}_<\vert _{z_0} $ is therefore a function of the fields  $\{ q_j^\prime \}$ at point $z_0$ and ${\cal Z}_>\vert _{z_0+\Delta z} $ is a function of $\{ q_j \}$. 

Let us now define the transfer integral operator $\hat T_{\Delta z}$ over the infinitesimally small distance $\Delta z$ through its action on an arbitrary 
function $\Psi $ of the field variables $q_j$ as 
\begin{equation} 
\hat T_{\Delta z} \Psi =C \int d^{\cal M}q^\prime \, 
e^{- \beta \, \Delta z \, h_\mathrm{fd} (\{ q_j^\prime \} , \, \{ q_j\}  ) } \Psi(\{ q_j^\prime \}) . 
\label{I.5} 
\end{equation} 
The constant $C=[\hbar ^2 /(\pi   \Delta z \, mk_BT)]^{{\cal M}/2}$ is introduced for the normalization of the Gaussian integrals and 
is not important for the rest of the derivation. 
Using the definition of $\hat T_{\Delta z}$ in Eq.~(\ref{I.2}) we can write
\begin{equation}
{\cal Z} = \int d^{\cal M}q \, {\cal Z}_>\vert _{z_0+\Delta z} \ \hat{T}_{\Delta z} \ {\cal Z}_<\vert _{z_0} . 
\label{I.4} 
\end{equation} 
Applying the transfer operator $N$ times, we obtain
\begin{equation}
{\cal Z} = \int{ d^{\cal M}q \, {\cal Z}_>\vert _{z_0+N\Delta z} \ \left(\hat{T}_{\Delta z}\right)^N \ {\cal Z}_<\vert _{z_0}} . 
\label{I.44} 
\end{equation} 
In general, we can define the transfer integral operator for a finite distance $|z^\prime -z|$ as the product of 
infinitely many transfer integral operators for infinitesimally small distances: 
\begin{equation} 
\hat T_{|z^\prime -z|}=\lim _{N\rightarrow \infty } \left( \hat T_{|z^\prime -z|/N}\right) ^N . 
\label{I.6}
\end{equation} 
Since we have assumed a homogeneous system, all transfer integral operators $\hat T_{\Delta z}$ for the same infinitesimally small distance $\Delta z$ are identical. 

To further simplify the equations we have to find the eigenfunctions of $\hat T_{\Delta z}$. In order to do so, we expand $\Psi (\{ q_j^\prime \}) $  in Eq.~(\ref{I.5}) around $\{ q_j \}$ up to second order 
in $q_j\prime -q_j$ and perform the integration. The result is $\hat T_{\Delta z} \Psi \approx (1-\Delta z \hat K)\Psi $ for $\Delta z\rightarrow 0$. As this is true for an arbitrary function $\Psi$, we can simply write
\begin{equation}
\hat T_{\Delta z}  \approx (1-\Delta z \hat K) \quad  \label{trans_K_connection}
\end{equation} 
for $\Delta z\rightarrow 0$. Here an auxiliary operator 
\begin{equation} 
\hat K=\sum _{j=1}^{\cal M} \left( -D\frac {\partial ^2}{\partial q_j^2}-\frac {\mu _j}{k_BT}q_j^2\right)  +
\frac {V (q_1, \, \dots \, ,\, q_{\cal M})}{k_BT} 
\label{K} 
\end{equation}
with 
\begin{equation}
D=\frac {mk_BT}{2\hbar ^2}       
\label{D} 
\end{equation}
was introduced.

The auxiliary operator (\ref{K}) has the structure of a quantum-mechanical Hamiltonian for a single particle in an external potential 
${U (q_1, \, \dots \, ,\, q_{\cal M})} =[{V (q_1, \, \dots \, ,\, q_{\cal M})}-\sum _{j=1}^{\cal M}\mu _jq_j^2]/(k_BT)$ in an 
${\cal M}$-dimensional space. Therefore, $\hat K$ has all the conventional properties of a single-particle Hamiltonian. 
In what follows, we consider only ``potentials" that yield a spectrum that is bounded from below. This is not the case for 
dynamically unstable systems, e.g., for an atomic Bose-gas with attractive interactions~\cite{Li-7in1D}. 
A stable system is characterized by an interaction that increases (or, at least, does not decrease) when the absolute value of any of the field components grows infinitely. 

To solve the eigenvalue problem 
\begin{equation} 
\hat K \Psi_\nu (q_1, \, \dots \, ,q_{\cal M} )=\kappa _\nu \Psi _\nu (q_1\, \dots \, ,q_{\cal M} ) 
\label{I.7} 
\end{equation} 
for a system characterized by an interaction that increases unlimitedly for $q_j\rightarrow \pm \infty $, we set the boundary conditions 
\begin{equation} 
\Psi _\nu \vert _{q_j\rightarrow \pm \infty } =0 , \qquad j=1,\, 2,\, \dots \, ,\, {\cal M} . 
\label{I.7bis} 
\end{equation} 
If the particular form of $V(q_1,\, \dots \, ,q_{\cal M})$ allows for the existence of a continuous spectrum, we can require, instead of 
Eq.~(\ref{I.7bis}) that the eigenfunctions are finite at ${q_j\rightarrow \pm \infty }$. 

The system of eigenfunctions 
is complete, orthogonal and normalized, like for a standard quantum-mechanical Hamiltonian problem. 
Note that the dimensionality of the eigenvalues $\kappa _\nu $ of the Hamiltonian-like operator $\hat K$ is inverse length and not energy.
The operator (\ref{K}) is  invariant with respect to  complex conjugation. Therefore the eigenfunctions of $\hat K$ can always be chosen to be real. In what follows we 
always assume  $\Psi _\nu ^*=\Psi _\nu $.

We assign the index $\nu =0$ to the lowest eigenvalue of $\hat K$ and, correspondingly, to the respective function. In what follows we 
refer to $\kappa _0 $ and $\Psi_0 $ as to the ``ground state" eigenvalue and eigenfunction, respectively. We assume that this 
ground state is not degenerate. If the operator $\hat K$ possesses a symmetry that makes the ground state degenerate, one can resort to a standard scheme of symmetry breaking~\cite{Bogoliubov1994mathematical}. 

Let us now go back to simplifying the expressions for the partition function Eq.~(\ref{I.44}). 
In the following, we will use the bra-ket notation 
$$\Psi_\nu (q_1, \, \dots \, ,q_{\cal M} )\equiv \langle q_1, \, \dots \, ,q_{\cal M} |\nu \rangle .$$
As one can easily see from Eq.~(\ref{trans_K_connection}), the eigenfunction of $\hat{K}$ are also eigenfunctions of $\hat T_{\Delta z}$. In the limit $ \Delta z\rightarrow 0$ we have
\begin{align}
\begin{split}
&\hat T_{\Delta z} \Psi_\nu   \approx (1-\Delta z \kappa_\nu) \Psi_\nu \approx \exp(- \Delta z \kappa_\nu )\Psi_\nu \label{trans_K_connection_eigfunc}.
\end{split}
\end{align} 
Using the spectral representation of $\hat T_{\Delta z}$ and the orthonormality of the eigenfunctions $|\nu \rangle$ we can simplify Eq.~(\ref{I.6}) for the transfer integral operator for a finite distance to
\begin{equation} 
\hat T_{|z^\prime -z|}=\sum _\nu  |\nu \rangle \langle \nu |\exp \left( -\kappa _\nu |z^\prime -z|\right)  .
\label{I.8}
\end{equation} 
One can see that for large distances $|z^\prime -z|$ the operator is dominated by the ground state $|0\rangle$.

By construction, ${\cal Z}_< \vert _{z_0} = \hat T _{z_0-z_{-1}} {\cal Z}_< \vert _{z_{-1}}$, where $z_{-1}<z_0$. 
Therefore, ${\cal Z}_< \vert _{z_0}$ is dominated by the ``ground state" eigenfunction, i.e.,
\begin{equation} 
{\cal Z}_< \vert _{z_0} = \mathrm{const} \, \Psi_0
\end{equation}
for a point $z_0$ far from the end points $z=\pm L/2$ . The same holds true for ${\cal Z}_> \vert _{z_0}$.
In the thermodynamic limit we can therefore write the partition function, up to an unimportant numerical factor, as 
\begin{equation}
{\cal Z} =  \ \langle 0| \hat{T}_{|z-z\prime|} |0\rangle  =  \exp \left( -\kappa_0 |z^\prime -z|\right). 
\label{part_c1_c2_exp}
\end{equation}

Let us now use the introduced quantities to calculate correlation functions. For simplicity, we will first focus on two-point functions. The generalization to a higher number of points can easily be done. 
In the quantum description, a two-point correlation function is 
\begin{equation} 
\langle \hat{\cal F}^{(1)}\vert _{z_1} \hat{\cal F}^{(2)}\vert _{z_2}\rangle = 
\mathrm{Tr}\, \left[ \hat{\cal F}^{(1)}\vert _{z_1} \hat{\cal F}^{(2)}\vert _{z_2}\hat \varrho \right] , 
\label{corr-q} 
\end{equation} 
where $\hat \varrho $ is the density matrix of the many-body quantum system. The operators $\hat{\cal F}^{(i)}\vert _{z_i} $ of the local observables are composed of normally ordered products of the field operators $\hat \psi ^\dag _j(z_i)$ and $\hat \psi (z_i)$ that create or, respectively, annihilate a particle at the point $z_i$.

In classical field approximation, the density matrix is replaced by the phase-space distribution function and the local observables become functions of the classical fields at the respective points. In thermal equilibrium, the correlation 
function of two observables $ {\cal F}^{(1)}\vert _{z_1}$ and ${\cal F}^{(2)}\vert _{z_2}$ at the points $z_1$ and $z_2$ is 
\begin{align}
\begin{split}
\langle {\cal F}^{(1)}\vert _{z_1} &{\cal F}^{(2)}\vert _{z_2}\rangle =  \\ &\frac 1{\cal Z} \int {\cal D}q_1 \, \dots \int {\cal D}q_{\cal M} \, e ^{-\beta H} {\cal F}^{(1)}\vert _{z_1} {\cal F}^{(2)}\vert _{z_2}.
\end{split}
\end{align}
For the sake of 
definiteness, we assume the ordering of the spatial points $z_2>z_1$ and can write
\begin{align} 
\begin{split}
\langle {\cal F}^{(1)}&\vert _{z_1} {\cal F}^{(2)}\vert _{z_2}\rangle  = \\
& \frac 1{\cal Z} 
\int d^{\cal M}q \ {\cal Z}_>{\vert }_{z_2} \, {\cal F}^{(2)}{\vert }_{z_2} \
\hat T_{z_2-z_1} \ {\cal F}^{(1)}{\vert }_{z_1} \, {\cal Z_<}{\vert }_{z_1} .  ~~~
\label{I.9} 
\end{split}
\end{align}  
Using Eqs.~(\ref{I.8}--\ref{part_c1_c2_exp}) we get
\begin{align} 
\begin{split}
\langle {\cal F}^{(1)}&\vert _{z_1} {\cal F}^{(2)}\vert _{z_2}\rangle \! = \left \{ \exp \left[ -\kappa_0 (z_2 -z_1)\right]\right\}^{-1} 
\\
&\times \sum _\nu    \langle 0| {\cal F}^{(2)}\! |\nu \rangle 
e^{-\kappa _\nu (z_2-z_1)}\langle \nu |{\cal F}^{(1)}\! |0\rangle
\\
 &= \! \sum _\nu \langle 0| {\cal F}^{(2)}\! |\nu \rangle 
e^{-(\kappa _\nu -\kappa _0)(z_2-z_1)}\langle \nu |{\cal F}^{(1)}\! |0\rangle  .     
\label{I.9bis} 
\end{split}
\end{align}
Here 
$$
\langle \nu ^\prime |{\cal F}^{(1)} |\nu \rangle =\int d^{\cal M}q\, \Psi _{\nu ^\prime } {\cal F}^{(1)}\Psi _\nu 
$$
is a standard quantum-mechanical matrix element (in a basis of real functions). 

Eq.~(\ref{I.9bis}) can be easily generalized to  an $l$-point correlation function that, in a general case, can contain 
$l$ different observables ${\cal F}^{(i)}$, $i=1,\, 2,\, \dots \, , \, l$: 
\begin{align} 
&
\langle {\cal F}^{(1)}\vert _{z_1} {\cal F}^{(2)}\vert _{z_2} \, \dots \, 
{\cal F}^{(l)}\vert_{z_l}\rangle = 
\sum _{\nu _1, \dots ,\nu_{l-1} } \langle 0|{\cal F}^{(l)}|\nu_{l-1}\rangle \dots  
~ \nonumber \\ & \quad \times 
\langle \nu_2|{\cal F}^{(2)}|\nu_{1}\rangle  \langle \nu_1|{\cal F}^{(1)}|{0}\rangle 
\prod _{i=1}^{l-1}e^{-(\kappa _{\nu_i} -\kappa _0)(z_{i+1}-z_i)} .
\label{FFF} \end{align} 
Here the spatial points are ordered as $z_l>\dots >z_2>z_1$.  

The case $l=1$ yields the thermal average of an arbitrary observable 
\begin{equation} 
\langle {\cal F}^{(i)}\rangle =\langle 0|{\cal F}^{(i)}|0\rangle . 
\label{I.10} 
\end{equation} 
An important particular case of Eq.~(\ref{I.10}) is the expression for the equilibrium distribution 
$W_{\mathrm{eq}}(q_1,\, \dots \, ,\, q_{\cal M})$ of the local values of the fields 
\begin{align} 
W_{\mathrm{eq}}(q_1,\, \dots \, ,\, q_{\cal M}) &= 
\int d^{\cal M}q^\prime \, |\Psi _0 (q_1^\prime ,\, \dots \, ,\, q_{\cal M}^\prime )| ^2 \nonumber \\ 
& \qquad \qquad \times \prod _{j=1}^{\cal M}\delta (q_j^\prime -q_j) \nonumber \\ 
&=|\Psi _0 (q_1 ,\, \dots \, ,\, q_{\cal M})| ^2 . \label{eq:W0}
\end{align} 

With this we finish the recollection of the basics of the transfer matrix formalism~\cite{Scalapino72,Krumhansl75,Currie80}. 
In the next Section we present our proof of its equivalence to a description based on a certain stochastic process. 

\section{From the transfer matrix to the stochastic process}
\label{from} 

Calculating correlation functions directly from Eq.~(\ref{FFF}) might be a challenging task, because of the need to know many eigenvalues and eigenvectors of $\hat K $. Therefore we have developed a method of numerical calculation of correlation functions in 1D that is 
based on the transfer matrix formalism but requires the knowledge of the ground state function $\Psi _0$ only. 

Assume that we know $\Psi _0$ and, hence, $W_\mathrm{eq}$. 
This task can be accomplished in many ways, from the diagonalization of $\hat K$ 
in a suitable basis or propagation of the corresponding Schr\"odinger-type equation in imaginary time to the variational method. 
Consider then a Fokker-Planck equation (FPE) 
\begin{align} 
& \frac {\partial W(q_1,\, \dots \, ,\, q_{\cal M})}{\partial z}= \sum _{j=1}^{\cal M} \left \{ 
D \frac {\partial ^2}{\partial q_j^2} W(q_1,\, \dots \, ,\, q_{\cal M}) \right. 
\nonumber \\ & \quad 
-\left.  \frac \partial {\partial q_j}[ A_{q_j}(q_1,\, \dots \, ,\, q_{\cal M})W(q_1,\, \dots \, ,\, q_{\cal M})]\right \} 
\label{FPE}
\end{align}
for the distribution function $W(q_1 ,\, \dots \, ,\, q_{\cal M})$ of the local values of the fields 
with the co-ordinate $z$ playing the role of time in the conventional version of the FPE. 
The diffusion coefficient $D$ is defined by Eq.~(\ref{D}).  
We then require $W_{\mathrm{eq}}$ defined by  Eq.~\eqref{eq:W0} to be the stationary solution of Eq.~(\ref{FPE}). This determines the drift coefficients 
\begin{align} 
A_{q_j}&=D \frac {\partial \ln W_\mathrm{eq} (q_1,\, \dots \, ,\, q_{\cal M})}{\partial q_j} \nonumber \\
&=2D \frac {\partial \ln |\Psi _0 (q_1,\, \dots \, ,\, q_{\cal M})|}{\partial q_j}. 
\label{A} 
\end{align} 
Note, that $\Psi _0$ possesses all the standard properties of a ground state function of a Hamiltonian problem, in particular, 
for all finite $q_j$'s it is non-zero and, hence, the coefficients $A_{q_j}$ have no singularities. This also means that we can always choose $\Psi _0$ to be real and positive, what we will assume for the rest of the paper.

In the following we will show that the correlation functions described by this FPE are identical to the ones we get from Eq.~(\ref{FFF}).
In order to do this we will use the well-known spectral method of solving the FPE~\cite{Risken89}. We introduce a new unknown function 
$\Xi (q_1^\prime ,\, \dots \, ,\, q_{\cal M}^\prime ,\, z)$ via the relation 
\begin{equation} 
W=\sqrt{W_\mathrm{eq}}\Xi =\Psi _0 \Xi . 
\label{II.1}
\end{equation} 
Substituting Eqs.~(\ref{A},~\ref{II.1}) into Eq.~(\ref{FPE}), we obtain 
\begin{equation} 
\frac {\partial \Xi }{\partial z} = \sum _{j=1}^{\cal M} \left( D \frac {\partial ^2}{\partial q_j^2} -\frac D{\Psi _0} 
\frac {\partial ^2\Psi _0}{\partial q_j^2} \right) \Xi .
\label{II.2} 
\end{equation} 
Eq.~(\ref{II.2}) 
can be  transformed  into 
\begin{equation} 
\frac {\partial \Xi }{\partial z} = - \hat K \Xi + \Xi \frac{1}{\Psi _0} \hat K \Psi _0. 
\label{II.3} 
\end{equation} 
By definition, $\hat K \Psi _0=\kappa _0\Psi _0$, where $\hat K$ is given in Eq.~(\ref{K}). With this we get 
\begin{equation} 
\frac {\partial \Xi }{\partial z} = -(\hat K -\kappa _0)\Xi . 
\label{II.3bis} 
\end{equation} 
The general solution of Eq.~(\ref{II.3bis}) is
\begin{equation}
	\Xi = \sum_\nu c_\nu \, \Psi _\nu (q_1,\, \dots \, ,\, q_{\cal M}) \,  e^{-(\kappa_\nu -\kappa _0)z} \label{fund_sol},
\end{equation}
where $c_\nu$ are the constant coefficients.

Let us now look at the  probability density $W_\mathrm{c}(\{q_j\},\, z_2|
\{q_j^\prime\} ,\, z_1)$ for $q_1,\, \dots \, ,\, q_{\cal M}$ at $z=z_2$ 
under the condition that the field values at $z=z_1<z_2$ are $q_1^\prime ,\, \dots \, ,\, q_{\cal M}^\prime $. 
This conditional probability density is a solution of Eq.~(\ref{FPE}) with the initial condition 
$$
W_\mathrm{c}(\{q_j\},\, z_1|\{q_j^\prime\} ,\, z_1)=\prod _{j=1}^{\cal M}\delta (q_j-q_j^\prime ). 
$$
Using the completeness of the set of eigenfunctions $\Psi _\nu $
$$
\prod _{j=1}^{\cal M}\delta (q_j-q_j^\prime )=\sum _\nu 
\Psi _\nu (\{q_j\}) \;
\Psi _\nu (\{q_j^\prime\} )  
$$
we find the particular values of the coefficients $c_\nu$ in Eq.~(\ref{fund_sol}). We obtain for $z_2>z_1$ 
\begin{align} 
\begin{split}
 W_\mathrm{c}&(\{q_j\} ,\, z_2| \{q^\prime_j\} ,\, z_1)=
\frac {\Psi _0(\{q_j\})}{\Psi _0(\{q^\prime_j\} )} 
 \\ &  \times \sum _\nu
{\Psi _\nu (\{q_j\} )}
 \ e^{-(\kappa _\nu -\kappa _0)(z_2-z_1)} \ 
{\Psi _\nu (\{q^\prime_j\})}   .
\label{II.4} 
\end{split}
\end{align}  

The equilibrium distribution of local values of the fields is given by the stationary solution of Eq.~(\ref{FPE}), 
which is $W_\mathrm{eq}$. The two-point correlation function can therefore be written as 
\begin{align}
\begin{split}
& \langle  {\cal F}^{(1)}\vert _{z_1} {\cal F}^{(2)}\vert  _{z_2}\rangle = \int d^{\cal M}q \, 
{\cal F}^{(2)}(\{q_j\} ) \\ & \times \int d^{\cal M}q^\prime \, 
W_\mathrm{c}(\{q_j\},\, z_2|
\{q_j^\prime\} ,\, z_1) \,
{\cal F}^{(1)}(\{q_j^\prime\}) \, W_\mathrm{eq}(\{q_j^\prime\} ).
~\label{II.5}
\end{split} 
\end{align} 
Using the solution Eq.~(\ref{II.4}) one can easily see the equivalence to Eq.~(\ref{I.9bis}). Convolving the conditional probability densities subsequently, one can show that 
the $l$-point correlation function for the stochastic process described by the FPE is given in the most general case 
by Eq.~(\ref{FFF}). We can therefore conclude that the transfer matrix formalism and the description by the FPE are equivalent, since the correlation functions following from either of these methods 
are identical. 

We recall 
the well-known equivalence of the FPE and the stochastic differential It\={o} equation~\cite{Risken89,GardinerBook,zinn2002quantum} 
\begin{equation} 
dq_j = A_{q_j}dz + \sqrt{2D}\, dX_j,
\label{Ito} 
\end{equation} 
where $dX_j$ are infinitesimally small, mutually uncorrelated, random terms obeying
 Gaussian statistics with zero mean and the variance
equal to $dz$: $\overline{dX_j}=0$, $\overline{dX_j dX_{j^\prime }}=\delta _{jj^\prime }dz$. Here the bar denotes averaging over the ensemble of realizations of the 
stochastic process. The initial values (say, at $z=0$) of the fields for each realization are obtained by (pseudo)random 
sampling their equilibrium distribution $W_\mathrm{eq}$. The subsequent numerical integration of Eq.~(\ref{Ito}) and 
averaging over many realizations yields the correlation functions. 

Eq.~(\ref{Ito}) is therefore  a generalization of our previous method  to simulate  
the classical thermal fluctuations in a system described by a quadratic Hamiltonian using the Ornstein-Uhlenbeck 
stochastic process~\cite{stimming2010fluctuations}
to the case of the arbitrary local interaction $V$. The main advantage of the stochastic \textit{ordinary} differential Eq.~\eqref{Ito} 
is that its numerical integration is much simpler and less resource-consuming than the integration of the 
\textit{partial} differential   Eq.~(\ref{FPE}) on a multidimensional grid. 
The main computational difficulty is now reduced  to the precise determination 
of $\Psi _0$. 
The determination of 
all other eigenfunctions and eigenvalues 
appearing in Eq.~(\ref{FFF}) is actually not necessary. 

\section{Tunnel-coupled quasi-condensates in 1D: The full model vs. the sine-Gordon model} 
\label{model}

We apply our method to the calculation of thermal phase and density fluctuations of two tunnel-coupled 1D quasi-condensates of 
ultracold bosonic atoms. The quasi-condensates in the right (\textit{R}) or in the left (\textit{L}) 
1D atomic waveguide are described in 
the mean-field approximation by complex classical fields $\psi _R\equiv q_1+iq_2$ and $\psi _L=q_3+iq_4$, respectively.
Alternatively, it is possible to express these complex fields $\psi _\varsigma =\sqrt{n_\varsigma }e^{i\theta _\varsigma }$, 
$\varsigma =R,\, L$, through the quasi-condensate atom-number densities $n_{R,L} $ and phases $\theta _{R,L} $~\cite{Mora03}. 
This system is described by Eq.~(\ref{hm1}) with the interaction term~\cite{Goldstein97,Whitlock03} 
\begin{align} 
V(q_1, q_2, q_3, q_4) = &\frac g2 \left[ \left(q_1^2+q_2^2\right)^2+\left(q_3^2+q_4^2\right)^2 \right] \nonumber \\ &-2\hbar J \left(q_1 q_3 + q_2 q_4\right), 
\label{VgJ}
\end{align}  
where $g > 0$ is the strength of the contact interaction of atoms in 1D and $J$ is the single-particle tunneling rate.
The tunneling provides exchange of atoms between the two waveguides, therefore atoms in both of them have the same 
chemical potential $\mu _R=\mu _L \equiv \mu $. The mean 1D atom-number density in each of the waveguides that 
corresponds to this chemical potential is denoted by $n_\mathrm{1D}$.  

It is convenient to parametrize the density variables as $n_\varsigma = r^2_\varsigma n_\mathrm{1D}$. Then we obtain 
\begin{equation}
\hat{K} =\frac 1{\lambda_T }\left[ \hat{\cal K}_R + \hat{\cal K}_L - 2 b \, r_R r_L \cos(\theta_R-\theta_L)\right] , \label{K_dens_phase}
\end{equation}
where
\begin{align}
\begin{split}
\hat{\cal K}_{\varsigma} = &-\left( \frac{\partial^2}{\partial r_\varsigma^2} + 
\frac{1}{r_\varsigma} \frac{\partial}{\partial r_\varsigma} + \frac{1}{r_\varsigma^2} \frac{\partial^2}{\partial \theta_\varsigma ^2} \right)  
\\ &+ \alpha \, r_\varsigma^2 \left[ r_\varsigma^2 - 2 \left( \tilde{\mu} - \frac{b}{2 \alpha} \right) \right] .
\label{KKo} 
\end{split}
\end{align}
The dimensionless parameters of the problem are 
\begin{equation}
\alpha = 
\frac {\lambda _T^2}{4\xi_\mathrm{h} ^2} , 
\qquad 
b =\frac { \lambda_T^2}{8 l_J^2},      \label{def.beta} 
\end{equation}
where $\xi_\mathrm{h} =\hbar /\sqrt{gn_\mathrm{1D} m}$ is the quasi-condensate healing length,  
$\lambda_T = 2 \hbar^2 n_\mathrm{1D}/(m k_B T)$ is the thermal coherence length and $l_J = \sqrt{\hbar/(4 m J)}$ 
is the typical length of the relative phase locking~\cite{stimming2010fluctuations,grivsins2013coherence}. 
The parameters $\alpha$ and $b$ can be understood as the ratio of the energies of the mean-field repulsion 
and of the tunnel coupling, respectively, to the kinetic energy of an atom localized at the length scale of the order of $\lambda _T$. 
Note that 
$\tilde \mu =\mu/(gn_\mathrm{1D} )$ is not a free parameter, but 
has to be chosen such that the average 1D density equals $n_\mathrm{1D}$ 
in both waveguides, i.e., $\langle r_L^2\rangle = \langle r_R^2\rangle =1$.
Therefore the eigenstates of $\hat{K}$  depend on $\alpha$ and $b$ only. Since $D = n_\mathrm{1D}/\lambda_T$, the solution of the It\={o} equation \eqref{Ito} also depends on the scaled distance $z/\lambda_T$.

In the following we will focus on discussing the relative phase fluctuations $\theta_-=\theta_R - \theta_L$, 
because they can be accessed experimentally through matter-wave interferometry~\cite{Schumm05}.
While it only makes sense to discuss the phase $\theta_-(z)$ modulo $2\pi$ for a single point, the unbound phase differences $\theta_-(z) - \theta_-(z')$ between two different points $z$ and $z'$ have a physical meaning. We obtain continuous phase profiles from the numerical samples of $\psi_{R,L}$ through phase unwrapping, i.e., by assuming that $\theta_-$ between neighbouring points on the numerical grid does not differ by more then $\pi$. The same procedure has been applied to experimental data in Ref.~\cite{Schweigler17}. 

We will compare the results for the two coupled quasi-condensates  to the predictions of the SG model  
\begin{align} 
\begin{split}
\label{eq:SG}
{H}_{\mathrm{SG}} =   \int _{-L/2}^{L/2}{d}z \Bigg[  gn_-^2 + & \frac{\hbar^2 n_\mathrm{1D}}{4m} 
\left(\frac {\partial {\theta_-}}{\partial z}\right) ^2  \\
& \quad \quad - 2 \hbar J n_\mathrm{1D} \cos{\theta_-} \Bigg] \, \mathrm{,}
\end{split}
\end{align} 
where $n_-=(n_R-n_L)/2$. 
The SG model was proposed as an effective model for the coupled quasi-condensates~\cite{Gritsev07}. 
Its validity in a certain parameter regime was recently confirmed experimentally~\cite{Schweigler17}.  
Since Eq.~\eqref{eq:SG} does not contain terms coupling $n_-$ to $\theta _-$, the relative density fluctuations can be integrated out 
and  the relative phase correlations are fully determined by the eigensystem of  
the auxiliary Hermitian operator for the  SG model~\cite{grivsins2013coherence}  
\begin{equation}
\hat{K}^{\mathrm{SG}} = \frac 1{\lambda_T }\left( - 2 \frac{\partial^2}{\partial \theta_-^2} - 2 b  \cos\theta_-\right) ,
\label{KSG} 
\end{equation}
which can be formally obtained from Eq.~\eqref{K_dens_phase} by setting $r_L = r_R \equiv 1$, $\partial /\partial r_{R,L}\equiv 0$, and 
$\partial ^2/\partial \theta _{R,L}^2 \equiv \partial ^2/\partial \theta _-^2$. Note that Eq.~(\ref{KSG}) does not contain the 
parameter $\alpha $, i.e., the equal-time phase correlations in the SG model at finite temperature do not depend on the 
atomic interaction strength. 

Due to the simpler nature of the model we can  obtain results directly from the transfer matrix formalism, without 
numerical implementation of Eq.~(\ref{Ito}). 
This will be the topic of the next Section~\ref{sGr}. The numerical results for the full model will be presented in Section~\ref{rid}.

\section{Results for the sine-Gordon model}
\label{sGr} 

In contrast to the previous work~\cite{grivsins2013coherence}, we consider the moments of the relative phase difference 
$\theta_- (z) - \theta_- (z^\prime ) $ itself 
and not the correlation function $\langle \exp (i [\theta_- (z) - \theta_- (z^\prime )]) \rangle$ for its imaginary exponent. The system is translationally invariant, therefore we can set 
$z^\prime =0$. We also assume $z>0$. 
Eq.~(\ref{KSG}) is invariant against inversion of the sign of $\theta _-$. Therefore only even moments 
$\langle [\theta _-(z)-\theta _-(0)]^k \rangle $, $k=2,\, 4,\, 6,\, \dots $, are non-zero. We express them as 
\begin{equation} 
\langle [\theta _-(z)-\theta _-(0)]^k \rangle =(-i)^k \lim _{\varepsilon \rightarrow 0} 
\frac {\partial ^k}{\partial \varepsilon ^k} {\cal C}(\varepsilon ,z),
\label{C.1} 
\end{equation}
where
\begin{equation} 
{\cal C}(\varepsilon ,z) = \langle \exp \{ i\varepsilon [\theta _-(z)-\theta _-(0)]\} \rangle . 
\label{C.1bis} 
\end{equation} 

Note that Eq.~\eqref{KSG} is equivalent to a Hamiltonian with a periodic potential. 
Bloch's theorem therefore tells us that the eigenfunctions must fulfill quasi-periodic boundary conditions, i.e., they must be Bloch waves.
However, when evaluating $\langle \exp [i\theta _-(z)-i\theta _-(0)]\rangle $ using Eq.~(\ref{I.9bis}), only the eigenfunctions satisfying periodic boundary conditions over $2\pi $ are relevant~\cite{grivsins2013coherence}.
In contrast, the direct evaluation of ${\cal C}(\varepsilon ,z)$ with $\varepsilon <1$ also requires knowledge of the eigenfunctions satisfying quasi-periodic boundary conditions over $2\pi $.
However, there is a trick that allows us to avoid this 
demanding calculation and use the $2\pi $-periodic eigenfunctions only.

We observe that applying Eq.~(\ref{I.9bis}) yields 
\begin{equation} 
{\cal C}(\varepsilon ,z) =\langle 0| e^{i\varepsilon \theta _-}
\exp [-(\hat K^\mathrm{SG}-\kappa _0)z]e^{-i\varepsilon \theta _-}|0\rangle . 
\label{C.2} 
\end{equation} 
Note that the lowest eigenvalue $\kappa _0$ corresponds to the state $\langle \theta _-|0\rangle $ that is periodic over $2\pi $. 
We now observe that $e^{i\varepsilon \theta _-}$ and its inverse, $e^{-i\varepsilon \theta _-}$, perform a unitary transformation of 
the operator $\exp [-(\hat K^\mathrm{SG}-\kappa _0)z]$, yielding 
\begin{align} 
{\cal C}(\varepsilon ,z) =&\langle 0| 
\exp \bigg{ \{ }-\bigg{[} -\frac 2{\lambda _T}\left( \frac{\partial ~}{\partial \theta_-} -i\varepsilon \right)^2  
\nonumber \\ &    
-\frac {2b}{\lambda _T}\cos \theta _- -\kappa _0\bigg{]} z\bigg{ \} } 
|0\rangle \nonumber \\  = & 
\exp \left( -\frac {2\varepsilon ^2z}{\lambda _T}\right) {\cal J}(\varepsilon ,z),   
\label{C.2bis} 
\end{align}
where 
\begin{equation} 
{\cal J}(\varepsilon ,z)=\langle 0| \exp \left[ -\left( \frac {4i\varepsilon }{\lambda _T}\frac{\partial ~}{\partial \theta_-}
+\hat K^\mathrm{SG}-\kappa _0\right) z \right] |0\rangle .  
\label{C.3}
\end{equation} 
The evaluation of the correlations via Eqs.~(\ref{C.1},~\ref{C.1bis}) 
is then reduced to expanding the exponential in Eq.~(\ref{C.3}), differentiating 
Eq.~\eqref{C.2bis} over $\varepsilon $ and  setting the limit $\varepsilon \rightarrow 0$. The resulting expression is an infinite sum 
of powers of $\hat K^\mathrm{SG}-\kappa _0$ and the derivative $\partial /\partial \theta _-$. Only the latter operator 
has off-diagonal matrix elements in the basis of eigenstates of $\hat K^\mathrm{SG}$. When we calculate the $k$th moment 
of the relative phase difference, the operator $\partial /\partial \theta _-$ appears in each term of the infinite sum 
maximum $k$ times. If it acts on an eigenfunction that is periodic over $2\pi $ then the resulting function is also periodic. 
Therefore, we avoid the necessity to calculate quasi-periodic eigenfunctions as the inner product of a periodic and a quasi-periodic functions vanishes. Assembling the series yields for the second and 
fourth moments 
\begin{widetext} 
\begin{align}
\left\langle [\theta _-(z)\! -\! \theta _-(0)]^2 \right\rangle &= 
\frac {4z}{\lambda _T} -  \left( \frac{4 z}{\lambda _T} \right)^2 
\sum _\nu | \wp _{\nu 0}| ^2 \, _1F_1 \big{(} 1;3;-(\kappa _\nu -\kappa _0)z \big{)} ,
\label{C.4} \\ 
\left\langle [\theta _-(z)\! -\! \theta _-(0)]^4 \right\rangle &= 
3 \, \left\langle [\theta _-(z)\! -\! \theta _-(0)]^2 \right\rangle ^2 + 
3  \left( \frac {4 z}{\lambda _T}\right) ^4 \Bigg{ \{  } 
8 \sum _{j_1=0}^\infty \sum _{j_2=0}^\infty \sum _{j_3=0}^\infty 
\frac {(-1)^{j_1+j_2+j_3}}{(j_1+j_2+j_3+4)!}
\sum _{\nu_1,\nu_2,\nu_3} \wp _{0\nu_1}\wp_{\nu_1\nu_2} \nonumber \\ & \times   
\wp_{\nu_2\nu_3}\wp_{\nu_3 0}
(\kappa _{\nu_1} -\kappa _0)^{j_1}(\kappa _{\nu_2} -\kappa _0)^{j_2}(\kappa _{\nu_3} -\kappa _0)^{j_3}z^{j_1+j_2+j_3} \!   
-\! \left| \sum _\nu | \wp _{\nu 0}| ^2 \, _1 F _1 \big{(} 1;3;-(\kappa _\nu -\kappa _0)z \big{)} \right| ^2\Bigg{ \}  } .  
\label{C.5}  
\end{align} 
\end{widetext}
Here 
$$
\wp_{\nu \nu^\prime } =\int _{-\pi}^\pi d\theta \, \Psi _\nu (\theta ) \frac {d \Psi _{\nu ^\prime }(\theta )}{d\theta } ,
$$
where $\Psi _\nu $ is the $\nu $th eigenfunction of the operator $\hat K^\mathrm{SG}$ satisfying periodic boundary conditions 
on the interval from $-\pi $ to $\pi $.  The set of $\Psi _\nu (\theta _-) $ can be found 
by the numerical diagonalization of $\hat K^\mathrm{SG}$ in the Fourier (i.e., cosine and sine) basis. Furthermore, we expressed the series 
$$
\sum _{j=0}^\infty \frac {2x^j}{(j+2)!}=\frac {2(e^x-1-x)}{x^2} =\, _1F_1(1;3;x)
$$
via the confluent hypergeometric function $_1F_1(a;c;x)$~\cite{Abramowitz}. 
In the next section we will present the results of the full model and compare them to Eqs.~(\ref{C.4},~\ref{C.5}).

\section{Results for the full model and discussion}
\label{rid} 

\subsection{The ground state} 

Finding the ground state of the operator $\hat K$ for the full model [Eqs.~(\ref{K_dens_phase},~\ref{KKo})] 
is a formidable task. However, the general structure of the operator \eqref{K} [and, hence, of Eq.~\eqref{K_dens_phase}]  that 
contains only local pairwise interactions admits for a solution. First of all, we notice that Eq.~\eqref{K_dens_phase} 
is invariant with respect to 
simultaneously shifting  both the angles $\theta_R $ and $\theta_L$ by the same value. Therefore, the (non-degenerate) ground state 
$\Psi _0$ must be independent of $\theta _+=\theta_R + \theta_L$. 

We search the ground state in the form 
\begin{align}
\Psi _0(r_R,r_L,\theta _-)=&\sum _{\ell =0}^{\ell _\mathrm{max}}\sum _{n_r,n_r^\prime =0} ^{n_{r\, \mathrm{max}}}
c_{\ell ,n_r,n_r^\prime }\Phi _{n_r \ell }(r_R) \Phi _{n_r^\prime  \ell }(r_L)\nonumber \\ & \times 
\frac {\cos (\ell \theta _-)}{\sqrt{\pi (1+\delta _{\ell 0})}}, 
\label{Psi0.a} 
\end{align} 
where $\Phi _{n_r \ell }(r_\varsigma )$, $\varsigma =R,\, L$, are the eigenfunctions of the operator 
\begin{align}
\begin{split}
\hat{\cal K}_{\varsigma}^{(\ell )} = &-\left( \frac{\partial^2}{\partial r_\varsigma^2} + 
\frac{1}{r_\varsigma} \frac{\partial}{\partial r_\varsigma}\right) + \frac{\ell ^2}{r_\varsigma^2}   
\\ &+ \alpha \ r_\varsigma^2 \left[ r_\varsigma^2 - 2 \left( \tilde{\mu} - \frac{b}{2 \alpha} \right) \right] , 
\label{KKl} 
\end{split}
\end{align}
labelled by the number $n_r$ of its nodes at $0 < r_\varsigma < \infty $.
Eq.~(\ref{KKl}) is obtained from Eq.~(\ref{KKo}) by replacing $\partial ^2 /\partial \theta _\varsigma ^2$ with $-\ell ^2$. 
The normalization condition for the (real) eigenfunctions of $\hat{\cal K}_{\varsigma}^{(\ell )} $ is 
$$
\int _0^\infty dr\, r \, \Phi _{n_r \ell }(r)\Phi _{n_r^\prime  \ell }(r)=\delta _{n_r^\prime  n_r}.  
$$

The most convenient basis to diagonalize $\hat{\cal K}_{\varsigma}^{(\ell )}$ is given by the radial eigenfunctions of a 
two-dimensional harmonic oscillator~\cite{Fluegge,Abramowitz}:
we expand $\Phi _{n_r \ell }(r_\varsigma )$ in the functions
$$
R_{n \ell} = \eta \, \sqrt{2n! /(n+\ell )!} \ (\eta r_\varsigma )^\ell  \exp (-\eta ^2r_\varsigma ^2/2) \, L_{n}^\ell (\eta ^2r_\varsigma ^2),
$$ 
with $n = 0,\,1, \, \dots, \, n_\mathrm{max}$. The limiting number $n_\mathrm{max}$ for the expansion of $\Phi _{n_r \ell }(r_\varsigma )$ should not be confused with $n_{r\, \mathrm{max}}$ in Eq.~\eqref{Psi0.a}. In the equation, $L_{n}^\ell $ is the generalized Laguerre polynomial with $\ell \geq 0$ and $\eta $ is a numerical scaling factor. 
Experience showed that the best convergence is attained for $\eta \approx 5$. 
The key advantage of this choice of basis is that the matrix elements 
of the operator $\hat{\cal K}_{\varsigma}^{(\ell )}$, consisting of the Laplacian in the polar co-ordinates and the powers $r_\varsigma ^2$, $r_\varsigma ^4$, can be found analytically. Analytic expressions for the matrix elements  of 
$ r_\varsigma \exp (\pm i\theta _\varsigma )$  also exist in this basis. All these matrix elements   needed for the diagonalization of the complete operator $\hat{K}$ 
can be easily computed by expressing them   through the rotated bosonic creation and annihilation operators for a two-dimenasional harmonic oscillator~\cite{HO2D1974,HO2D2014}. 

The ground state $\Psi _0$ is found from the diagonalization of $\hat K$ as the lowest-eigenvalue solution. 
The diagonalization happens in the basis of the constituent functions given in Eq.~\eqref{Psi0.a}.
Theoretically, one would have to diagonalize an infinitely large matrix, practically one only has to diagonalize a submatrix which size is determined by the values for ${\ell _\mathrm{max}}$ and $n_{r\, \mathrm{max}}$. 
They are chosen to be large enough to ensure convergence.
For the results presented in this paper, we chose ${\ell _\mathrm{max}} = 20$ and $n_{r\, \mathrm{max}} = 17$. Convergence was checked by comparing to the results for ${\ell _\mathrm{max}} = 13$ and $n_{r\, \mathrm{max}} = 10$.
We begin the diagonalization with the initial 
guess $\tilde \mu = 1 - b/(2\alpha)$ for the chemical potential. After obtaining $\Psi _0$, we can compute $\langle r^2_{R,L}\rangle$. For symmetry reasons, it does not matter whether we look at R or L. By assumption, 
this value $\langle r^2\rangle $ must be equal to 1, but the first try will yield some value $\langle r^2\rangle \neq 1$. Therefore, we rescale $\tilde \mu $ to the value 
$\tilde \mu + 1 - \langle r^2\rangle $ and repeat the procedure until we obtain $\langle r^2\rangle =1$ with the required precision. For the data presented in this paper, we required $|\langle r^2\rangle -1|<10^{-4}$. 

Note that a quick evaluation of $\Psi_0$ [Eq.~\eqref{Psi0.a}] is paramount for an efficient solution of Eq.~\eqref{Ito}. Pre-calculating $\Psi_0$ on a three dimensional grid is not an option due to the large number of grid-points that would be required. Calculating everything from scratch is also inefficient as the functions $\Phi _{n_r \ell }(r_\varsigma )$ themselves consist of sums with many terms. The best option therefore seems to be to pre-calculate each of the functions $\Phi _{n_r \ell }(r_\varsigma )$ and also $ \cos (\ell \theta _-)$ on the respective one-dimensional grids, then evaluating the sum in Eq.~\eqref{Psi0.a} every time we need a certain value $\Psi _0(r_R,r_L,\theta _-)$.
The grids used for the data presented in this paper had 1024 points. Convergence was checked by comparison with the results for only 512 grid-points. To save computational time we consider only the terms with $c_{\ell ,n_r,n_r^\prime } > 10^{-5} $ in the sum Eq.~\eqref{Psi0.a}. Again we checked for convergence by comparing to the results for a cutoff of only $10^{-3}$. 

\subsection{Correlation functions} 

After having obtained the ground state, we can integrate Eq.~(\ref{Ito}) by the forward Euler method, using a pseudo-random generator to simulate the random terms.
The step size in $z$-direction was chosen as $\delta z/\lambda_T = 1/4000$. We checked for convergence by comparing to $\delta z/\lambda_T = 1/1000$. The presented results have been calculated from $1.2 \times 10^5$ numerical realizations.

Despite having obtained the ground state $\Psi_0$ in polar coordinates, we perform the random process in Cartesian coordinates $q_j$ to avoid numerical pitfalls.
The derivatives of $\ln(W_\mathrm{eq})$ needed for the drift coefficients $A_{q_j}$ [Eq.~\eqref{A}]  are calculated as finite differences in polar coordinates, then transformed to Cartesian coordinates in the usual way. 
Calculating for example the derivative in $r_L$, we first find the two points on the radial grid that are closest to the actual value of $r_L$. For the values of the two remaining variables we accept the grid points closest to the actual value of $r_R$ and $\theta_-$, respectively. We then evaluate $\ln(W_\mathrm{eq})$ on this two points of the three parameter grid which differ only in $r_L$, take the difference and divide by the step of the radial grid.   

We will compare the results to the predictions of the SG model with the rescaled parameters $\tilde{\lambda}_T =\lambda_T / \langle 1/r_\varsigma^2 \rangle _\mathrm{reg} $ and $\tilde{b} =b \ \langle r_R r_L \rangle / \langle 1/r_\varsigma^2 \rangle _\mathrm{reg} $. Here $\langle 1/r_\varsigma^2 \rangle _\mathrm{reg} $ 
represents the regularized mean inverse density (in dimensionless units), for symmetry reasons the expectation value is the same for $\varsigma = L, R$. 
The regularization is necessary, because otherwise
the $\ell =0$ component of the ground state would yield a logarithmic divergence of the integral.  
Different regularizations have been tested and all yielded very close results. 
In the end we chose to simply exclude a very small region around $r_\varsigma = 0$.

The reason for the rescaling of the parameters becomes clear when looking at the operator $\hat{K}$ given in Eqs.~(\ref{K_dens_phase},~\ref{KKo}). 
We average it over the ground state fluctuations of $r_{R,L}$  (the need for regularization of $1/r_\varsigma ^2$ should be kept in mind). This averaging 
yields an operator 
\begin{align}
\begin{split}
& \hat{K}^{\mathrm{SG}}_\mathrm{rescaled} =  \quad \frac 1{\lambda_T }   \left\langle  \frac 1 {r_\varsigma^2}  \right\rangle _\mathrm{reg}  \\ & \quad \quad \times\left( - 2   \frac{\partial^2}{\partial \theta_-^2} - 2 b \langle r_R r_L \rangle  \left\langle \frac 1 {r_\varsigma^2} \right\rangle _\mathrm{reg}^{-1} \cos\theta_-\right) .
\label{KSGreg}
\end{split}
\end{align}
The rescaling of $\lambda_T$ and $b$ can be directly read from the comparison of Eqs.~(\ref{KSG}) and (\ref{KSGreg}).  

\begin{figure}
	\centering
	\includegraphics[width=\linewidth]{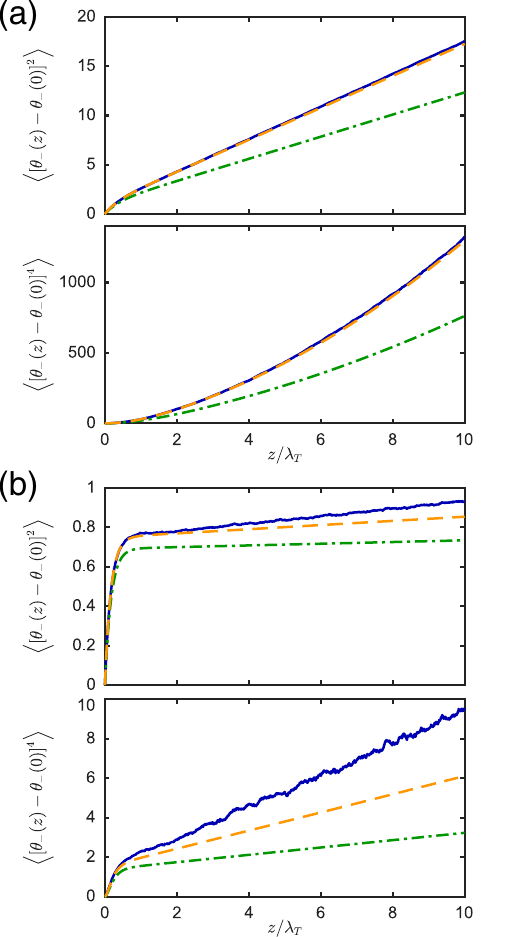}
	\caption{(Color online.) Second and fourth moment of the relative phase difference between two points along the 1D direction $z$. Results for $\alpha = 100$ and {(a)} $b = 1$, {(b)} $b = 5$. {(a)} In the intermediate phase-locking regime ($\left\langle \cos(\theta_-) \right\rangle = 0.58 $) we  observe good agreement between the two coupled 1D quasi-condensates (solid blue lines) and the sine-Gordon model with the rescaled parameters (dashed orange lines). Clear deviations from the  the sine-Gordon model without rescaling of the parameters (green dash-dotted lines) are visible. {(b)} For strong phase locking ($\left\langle \cos(\theta_-) \right\rangle = 0.83 $) we get clear deviations also for the rescaled sine-Gordon theory.}
	\label{fig:fig1}
\end{figure}

\begin{figure}
	\centering
	\includegraphics[width=\linewidth]{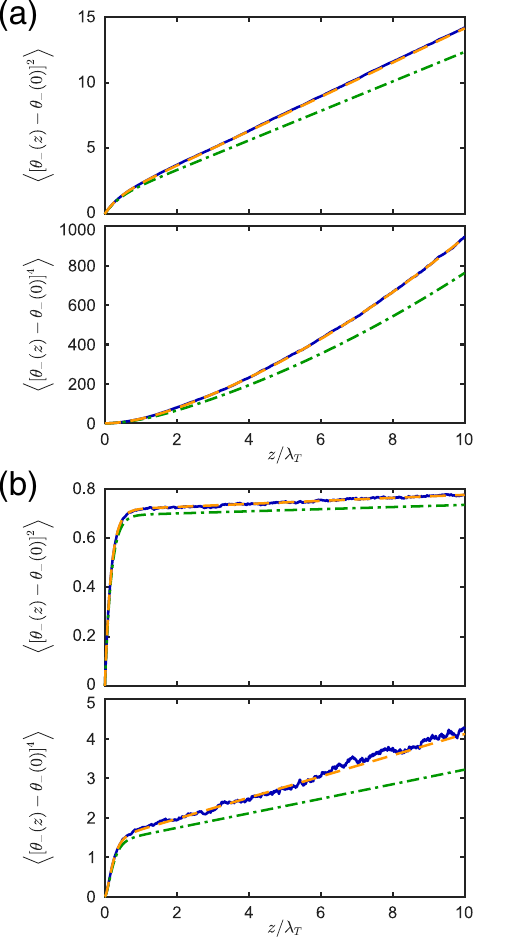}
	\caption{(Color online.) The same as in Fig.~\ref{fig:fig1}, but for $\alpha = 500$. In contrast to the case of $\alpha = 100$, there are no  visible deviations from the rescaled sine-Gordon theory, even for strong phase locking.}
	\label{fig:fig1_a}
\end{figure}

Fig.~\ref{fig:fig1} shows the results for $\alpha=100$.
For $b = 1$ (subfigure (a)), which corresponds to intermediate phase locking, one sees good agreement between the results for the full calculation and the rescaled SG model. The same is true for small phase locking (not shown). For stronger phase-locking $b = 5$  deviations are clearly visible [Fig.~\ref{fig:fig1}(b)]. For higher values of $\alpha$ (lower temperatures or higher densities) the agreement between the full theory and the rescaled SG model holds even for strong phase locking (see Fig.~\ref{fig:fig1_a} for $\alpha = 500$). One can understand the different behavior for the different values of $\alpha$ from the amount of density fluctuations being present in the system. The higher $\alpha $ (i.e., the more pronounced the 
effect of interatomic repulsion), the more suppressed are the density fluctuations. The accuracy of the 
SG description is thus increased. 
This explains the good agreement of the experimental data of Ref.~\cite{Schweigler17} with the SG model. The value of $\alpha$ there is rather high ($\alpha \approx 600$). Repeating the measurements for $\alpha = 100$ would be a challenging task due to the finite resolution of the imaging system. 

\begin{figure}
	\centering
	\includegraphics[width=\linewidth]{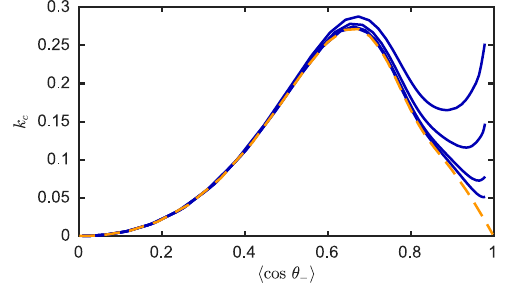}
	\caption{(Color online.) Circular kurtosis $k_c$ as defined in Eq.~\eqref{circ_kurt}. The solid blue lines represent the results for the coupled quasi-condensates, the different lines represent, from top to bottom, $\alpha = 100, 200, 500, 1000$. The dashed orange line represents the sine-Gordon prediction.}
	\label{fig:fig2}
\end{figure}

Note that, for the parameters in Fig.~\ref{fig:fig1}(b) ($\alpha = 100$ and $b = 5$),  it is not possible to achieve agreement between the full model and the SG theory by using a different rescaling. One can best see this from single-point expectation values calculated from $W_{\mathrm{eq}}$ \eqref{eq:W0}. They only depend on $b$ for the SG model and on $\alpha$ and $b$ for the coupled quasi-condensates.
We analyze  the circular kurtosis~\cite{fisher93}
\begin{equation}
k_c = \frac{\left\langle \cos(2 \theta_-) \right\rangle - \left\langle \cos  \theta_- \right\rangle^4}{(1 - \left\langle \cos \theta_- \right\rangle)^2}, \label{circ_kurt}
\end{equation}
which is a measure for the non-Gaussianity of the underlying distribution of $\theta_-$.
Fig.~\ref{fig:fig2} shows $k_c$  as a function of $\left\langle \cos \theta_- \right\rangle$. One can see the deviation of the exact results from the predictions of the SG model for $\langle \cos \theta _-\rangle \approx 1$. 
Again we see that the deviation from SG theory is bigger for smaller values of $\alpha$, i.e. for higher temperatures or lower densities. 

\begin{figure}
	\centering
	\includegraphics[width=\linewidth]{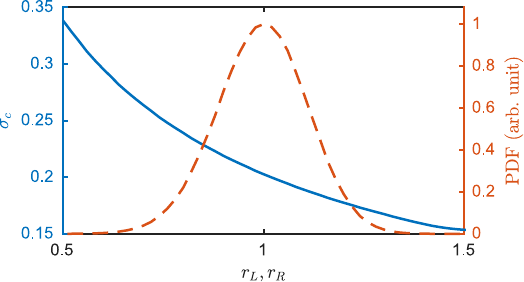}
	\caption{(Color online.) Circular standard deviation $\sigma_c = \sqrt{-2\ln\langle \cos \, \theta_- \rangle}$~\cite{fisher93} for the distribution of $\theta_-$ depending on the value $r_L = r_R$ (solid blue line, left vertical axis). The probability density function (PDF) integrated over $\theta_-$ for the different points $r_L = r_R$ is given in arbitrary units by the dashed red line. The corresponding  vertical axis is on the right.
	The results for $\alpha = 100$ and $b = 312.5$ are shown.
	}
	\label{fig:rdepsigma}
\end{figure}

Note that the non-Gaussianity for intermediate phase-locking (intermediate values of  $\left\langle \cos \theta_- \right\rangle$) and strong phase-locking ($\left\langle \cos \theta_- \right\rangle \approx  1$) has different physical origins. For intermediate phase-locking, $W_\mathrm{eq}(r_R,r_L,\theta_-)$ as a function of $\theta _-$ for fixed $r_R, r_L$ is non-Gaussian in the relevant range of $r_R$ and $r_L$ (close to 1). For strong phase-locking this is not the case any more. The distribution of $\theta_-$ for different points $r_R$, $r_L$ is approximately Gaussian, with the variance depending on $r_R$, $r_L$ (see Fig.~\ref{fig:rdepsigma}). Therefore, averaging over different points leads to an overall distribution for $\theta _-$ which is non-Gaussian.

\section{Conclusion}
\label{conclusion}

To summarize, we have developed a versatile method for calculating thermal expectation values for non-relativistic 1D bosonic systems. To be treatable by our method, a 1D system should satisfy the following criteria. First of all, it must be at thermal equilibrium. Secondly, the system must be dominated by classical thermal fluctuations; the quantum fluctuations should be negligible compared to them, which is often the case in weakly interacting atomic quasicondensates~\cite{stimming2010fluctuations}. Of course,  strongly interacting quantum systems (such as the Tonks-Girardeau gas) cannot be modeled by our method. The third criterion requires the dynamical stability of the system, the thermodynamic limit should exist, which is the case for atomic gases with repulsive interactions. The last condition is the locality of interactions This is not too restrictive, since short-range interactions are well approximated by a contact potential; tunnel coupling between waveguides and coherent electromagnetically driven transitions between spin states also comply with this restriction.

We applied the method to the case of two tunnel-coupled 1D quasi-condensates and compared the results to the predictions of the simpler sine-Gordon model for which we obtained analytical results. We identified the cases when the two descriptions agree and when their predictions differ. 

Our non-perturbative method is based on the stochastic ordinary  differential equation (It\={o} equation). 
The main advantage of the presented method is its computational efficiency. Calculating the $1.2 \times 10^5$ realizations used for Fig.~\ref{fig:fig1} and \ref{fig:fig1_a} takes around 2 hours on a desktop computer, which is at least by an order of magnitude shorter  than what more traditional methods like stochastic Gross-Pitaevskii (SGPE)~\cite{Blakie08} would need. 
Moreover, we should mention the robustness of our method in the presence of (quasi)topological excitation. Such excitations often comprise a problem when using methods based on the evolution in presence of a noise term (SGPE) or some sort of the Metropolis-Hastings algorithm~\cite{Grisins13}. We therefore believe that our method will find its application in a broad research area.
The ultimate goal of our work is to provide a method that could serve as a ``work horse'' in an experimental
laboratory, yielding reliable results with minimum resources in terms of both computer memory and computation time. 

\begin{acknowledgments}
The authors thank S. Erne, V. Kasper, and J. Schmiedmayer for helpful discussions. We acknowledge financial
support by the by the Wiener Wissenschafts und Technologie Fonds (WWTF) 	
via the grant MA16-066  
and by the EU via the ERC advanced grant QuantumRelax (GA 320975). This work was also supported by the Austrian Science Fund (FWF) via
the project P~25329-N27 (S.B., I.M.), the SFB ISOQUANT No. I 3010-N27, and the Doctoral Programmes W~1245-N25 ``Dissipation und Dispersion in nichtlinearen partiellen Differentialgleichungen" 
(S.B.) and W~1210-N25 CoQuS (T.S.). 
\end{acknowledgments}

\bibliography{biblioA1}

\end{document}